\documentclass{PoS}

\title{Direct detection data and possible hints for low-mass WIMPs }

\ShortTitle{Direct detection data and possible hints for low-mass WIMPs }

\author{\speaker{Thomas Schwetz}\\
        Max-Planck-Institute for Nuclear Physics,
        PO Box 103980, 69029 Heidelberg, Germany\\
        E-mail: \email{schwetz AT mpi-hd.mpg.de}}

\abstract{Possible hints for WIMP dark matter with mass around 10 GeV coming
  from the DAMA, CoGeNT, and maybe also CRESST experiments are presented, and
  confronted with constraints from CDMS and XENON data. Focusing on
  spin-independent (SI) WIMP--nucleus interactions, I elaborate on the
  difficulties to make the hints consistent with each other and to evade the
  constraints, mentioning energy scale uncertainties, quenching and
  light-yield factors, as well as uncertainties on halo properties. In the
  present situation it seems hard to reconcile all data within the SI
  framework, which suggests that if the experimental anomalies were indeed
  due to dark matter a more exotic mechanism (to be identified) had to be at
  work.}

\FullConference{Identification of Dark Matter 2010\\
		 July 26 - 30 2010\\
		 University of Montpellier 2, Montpellier, France}

\begin{document}

\begin{figure}[!h]
  \centering
  \includegraphics[width=0.55\textwidth]{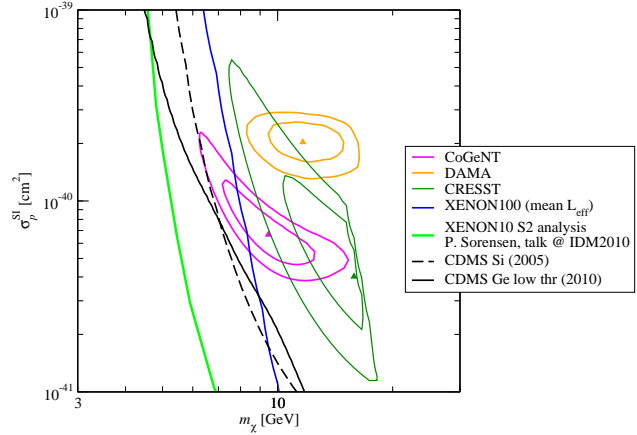}
  \caption{Hints from DAMA, CoGeNT, and CRESST at 90\% CL and $3\sigma$
  compared to the constraints from CDMS and XENON at 90\% CL for elastic
  spin-independent scattering.
  \label{fig:summary}}
\end{figure}

The dark matter direct detection experiments DAMA, CoGeNT, and maybe
also CRESST report some anomalies which can be interpreted in
terms of spin-independent (SI) scattering of WIMP dark matter
particles with a mass around 10~GeV and a WIMP--nucleon scattering
cross section of order $10^{-40} {\rm cm}^2$. This interpretation is
challenged by constraints mainly from the CDMS and XENON
experiments. The situation is summarized in Fig.~\ref{fig:summary}.

{\bf The hints.} The DAMA experiment reports evidence at about $8.9\sigma$ for an
annual modulation of their scintillation signal from NaI
crystals in the low energy region of the energy window, between 2 and
6~keVee~\cite{dama}. The phase of the modulation is in agreement with
the expectation from WIMP scattering due to the motion of the earth
around the sun. Following \cite{Bozorgnia:2010xy}, we assume that
channeling is negligible for nuclear recoils, which implies that
quenching has to be taken into account, i.e.\ only a certain fraction
of the recoil energy is deposited in scintillation light. The default
values are $q_{\rm Na} = 0.3$ and $q_{\rm I} = 0.09$ for sodium and
iodine recoils, respectively. The DAMA signal can be explained in
terms of SI scattering on either Na or I. However, the scattering on
iodine requires WIMP masses of order 70~GeV and cross sections
excluded by CDMS and XENON by more than two orders of
magnitude. Therefore, we focus here on scattering on Na, which, due to
its smaller nuclear mass is sensitive to lighter WIMPs, in the 10 GeV
region. The corresponding parameter region is shown by the orange
contours in Fig.~\ref{fig:summary}.

CoGeNT is a Ge detector with a very low threshold of 0.4~keVee and excellent
energy resolution. In \cite{cogent} they report an event excess between the
threshold and 3.2~keVee with an exponential shape, which cannot be explained
by known background sources and has a shape consistent with a signal from
WIMPs with a mass around 10 GeV. The CoGeNT region shown in
Fig.~\ref{fig:summary} has been obtained under the assumption that in the
signal region only identified peaks and a flat background are present, apart
from the WIMP signal. In particular, it has been assumed that there is no
background component with exponential shape. 

The CRESST-II experiment searches for WIMP recoils in a CaWO$_4$
target~\cite{cresst}. Using the relative signal in light and phonons
it is possible to distinguish nuclear recoil events from W and O as
well as $\alpha$ or $\gamma$ background events. At this conference an
unexplained excess of events in the O band has been
reported~\cite{cresst}. A total of 32 events has been observed with an
expected background of $8.4\pm 1.4$ events, mainly from $\alpha$'s
with a small contribution from neutrons and $\gamma$'s, see
\cite{cresst2} for a recent update confirming the excess. We have
performed a rough estimate of the region in WIMP mass and SI cross
section which could account for the signal by using the information
given in~\cite{cresst}. The region shown in Fig.~\ref{fig:summary}
assumes 400~kg~d exposure at 100\% efficiency and uses the
distribution of signal as well as background events within the various
CRESST detectors. The region is cut off for $m_\chi \gtrsim 20$~GeV by
requiring that the signal is consistent with the bound from the W band
given in~\cite{cresst} for the inelastic scattering scenario. While
this result is highly speculative and has to await confirmation from
the CRESST collaboration it is intriguing that the region ends up in
the same ballpark as the DAMA and CoGeNT hints. Given the still
unconfirmed nature of the CRESST signal we focus in the following on
the DAMA and CoGeNT results.

\begin{figure}
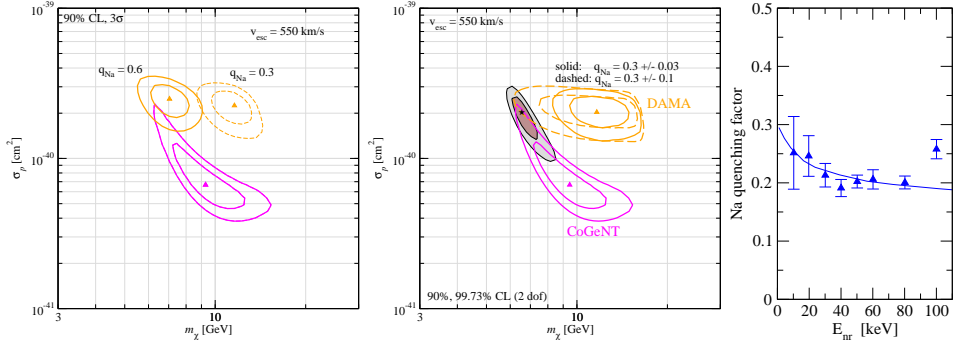

    \centering
    \includegraphics[height=0.3\textwidth]{cogent-dama.eps}
    \includegraphics[height=0.3\textwidth]{dama-vs-cogent.eps}
    \includegraphics[height=0.3\textwidth]{q.Na.eps}
    \caption{Left: DAMA region compared to CoGeNT for two different
      assumptions on the Na quenching factor, $q_{\rm Na} = 0.3$
      (dashed) and $q_{\rm Na} = 0.6$ (solid). Middle: DAMA regions
      marginalizing over the uncertainty on $q_{\rm Na}$ for two
      assumptions on the error. The shaded region corresponds to a
      combined DAMA/CoGeNT fit for $q_{\rm Na} = 0.3\pm 0.1$. Right:
      measurement of $q_{\rm Na}$ from \cite{Chagani:2008in}. 
      \label{fig:qNa}}
\end{figure}

From Fig.~\ref{fig:summary} it follows that although DAMA and CoGeNT
indicate similar WIMP masses, they require cross sections different by a
factor 2 to 5. In \cite{Hooper:2010uy} it has been suggested that
uncertainties in the quenching factor for Na could reconcile the two
results.  Fig.~\ref{fig:qNa} (left) shows that for larger $q_{\rm Na}$ the
DAMA region shifts to smaller WIMP masses which potentially could make it
consistent with the CoGeNT region. The DAMA regions in the middle panel have
been obtained by assuming a central value of $q_{\rm Na} = 0.3$ but allowing
to float it in the fit by adding a Gaussian penalty function to the $\chi^2$
assuming an error of $\pm0.03$ (solid contours) or $\pm 0.1$ (dashed
contours). In the latter case marginal overlap is found with CoGeNT which
might allow a combined fit (shaded region). The minimum $\chi^2$ of the
combined fit (assuming $q_{\rm Na} = 0.3\pm0.1$) is $\chi^2_{\rm comb} =
75/(68-4)$ corresponding to a probability of 1.6\%. In contrast, if each
experiment is fitted separately very good fits are obtained: $\chi^2_{\rm
DAMA} = 8.2/(12-2)$, (61\%) and $\chi^2_{\rm CoGeNT} = 46/(56-4)$, (71\%).
This indicates that sever tension remains in the combined fit. In the right
panel we show a recent measurment of $q_{\rm Na}$ from
\cite{Chagani:2008in}. These data suggest even a slightly lower value of
$q_{\rm Na}$ in the low energy region relevant for DAMA ($E_{nr} \simeq
10$~keV) of about $q_{\rm Na} \approx 0.25 \pm 0.05$. From this result it
seems unlikely that the uncertaintly on $q_{\rm Na}$ alone can make DAMA and
CoGeNT consistent with each other. Note also that shifting the energy scale
for CoGeNT due to uncertainties in the Ge quenching factors at these low
energies could move the region towards DAMA. However, background peaks in
the signal region at known possitions can be used to calibrate the CoGeNT
energy scale.

\begin{figure}
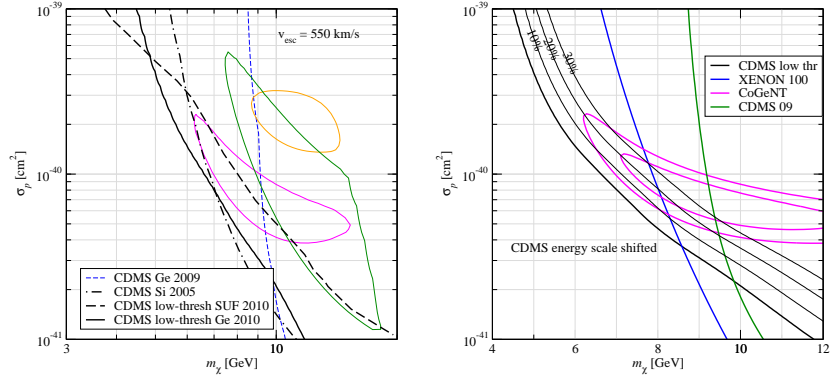

  \centering
  \includegraphics[height=0.33\textwidth]{cdms-all.eps} \quad
  \includegraphics[height=0.33\textwidth]{cdms-low-thr-Escale.eps}
  \caption{Left: constraints from different CDMS analyses: Ge 2009
    \cite{Ahmed:2009zw}, Si 2005 \cite{Akerib:2005kh}, low-thresh SUF
    2010 \cite{Akerib:2010pv}, and low-thresh Ge 2010
    \cite{Ahmed:2010wy}. Right: effect of a wrongly calibrated energy
    scale in the low-threshold Ge 2010 analysis. The thin black curves
    correspond to an energy scale shifted by 10\%, 20\%, and 30\%. \label{fig:cdms}}
\end{figure}

{\bf Constraints.} The non-observation of a significant excess of events in the CDMS
experiment \cite{Ahmed:2009zw} is a challenge for the dark matter
interpretation of these anomalies in terms of SI interactions.
Fig.~\ref{fig:cdms} (left) shows the bounds from CDMS coming from 
different analyses. The ``standard'' CDMS result labeled ``CDMS Ge 2009''
is based on a 10~keV threshold applying usual nuclear recoil
selection cuts, yielding 2 candidate events over an expected background of
$0.8\pm0.22$ \cite{Ahmed:2009zw}. Other analyses are less sensitive in the
more conventional WIMP region $m_\chi \gtrsim 50$~GeV, but can provide
stronger constraints on low-mass WIMPs. The ``Si 2005'' analysis
\cite{Akerib:2005kh} is based on 12~kg~d silicon data with a 7~keV
threshold. Thanks to the lower mass of Si compared to Ge and the slightly
lower threshold these data essentially exclude the CoGeNT/DAMA region. This
result is enforced by the recent ``low-threshold Ge 2010'' analysis
\cite{Ahmed:2010wy}, where cuts on nuclear recoil/electron event
discrimination are relaxed accepting some background but allowing to lower
the threshold to 2~keV. 

An important issue in interpreting such bounds is the calibration of the
energy scale of the detector, since the limit is dominated by the energy
threshold, where the largest signal is expected. Fig.~\ref{fig:cdms} (right)
illustrates how a hypothetical error in the energy scale calibration would
affect the limit from the low-threshold Ge 2010 analysis. It can be seen
that only a major shift in the energy scale of order 30\% can sufficiently
relax the bound. To the author it seems unlikely that such a large
mis-calibration happened, especially since the predicted/extrapolated
background matches well the observed event spectrum, see Fig.~1 of
\cite{Ahmed:2010wy}.

\begin{figure}
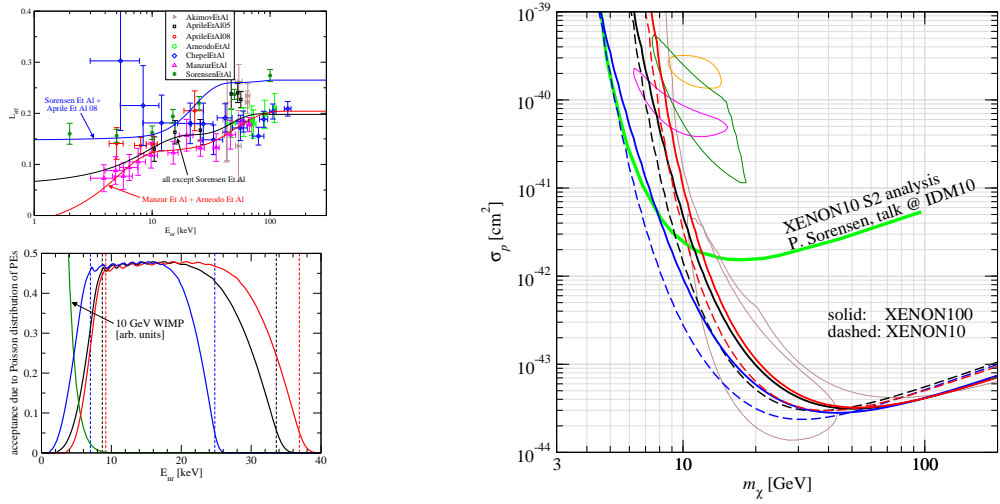

\centering
  \begin{tabular}{cc}
  \begin{minipage}{0.4\textwidth}
  \begin{tabular}{c}
    \includegraphics[width=0.7\textwidth]{Leff-data.eps}\\
    \includegraphics[width=0.7\textwidth]{poisson-resolution2.eps}
  \end{tabular}
  \end{minipage}
&
  \begin{minipage}{0.4\textwidth}
  \includegraphics[height=1.1\textwidth]{xenon-comp-S2.eps}
  \end{minipage}
  \end{tabular}
  \caption{Upper left: compilation of data on $L_{\rm eff}$ and three
      representative fits to part of the data. Lower left: the acceptance
      window in XENON100 for the three examples of $L_{\rm eff}$ compared to
      the signal expected from a 10~GeV WIMP. Right: exclusion curves from
      XENON10 (dashed) and XENON100 (solid) for the three examples of
      $L_{\rm eff}$. The color of the curves in the three panels indicate
      the corresponding choice for $L_{\rm eff}$. In the right panel also
      the constraint from an S2-only analysis of XENON10
      data~\cite{sorensen} is shown. \label{fig:xenon}}
\end{figure}

Let us now move to the results from the XENON10~\cite{xenon10} and
XENON100~\cite{Aprile:2010um} experiments, which also provide serious
constraints in the region of interest. Using a coincidence in signals from
scintillation (S1) and ionization (S2) an efficient nuclear recoil
identification is possible.  The energy scale for nuclear recoils is set by
the S1 signal, which has to be translated into nuclear recoil energy
$E_{nr}$ with the help of the light-yield function $L_{\rm eff}(E_{nr})$.
Measurments of $L_{\rm eff}$ are shown in the upper left panel of
Fig.~\ref{fig:xenon}. For the low-mass region especially the low energy
region of $L_{\rm eff}$ is important where data are scarce and partially
inconsistent. Therefore we adopt three representative curves shown in the
figure. As illustrated in the lower left panel the behaviour of $L_{\rm
eff}$ is crucial to constrain low mass WIMPs, since the effective acceptance
window depends on this choice, see also~\cite{savage}. The corresponding
limits are shown in the right panel of Fig.~\ref{fig:xenon}. It is clear
that if $L_{\rm eff}$ is not too low in the low energy region XENON10/100
results provide severe constraints on the CoGeNT/DAMA region. A recent
work \cite{Bezrukov:2010qa} exploring the correlation of
scintillation and ionization signals in xenon suggests $L_{\rm eff}$ values
somewhere between the black and the blue curves in Fig.~\ref{fig:xenon},
enforcing the xenon constraints.

At this conference a preliminary analysis of XENON10 data has been
presented, based only on the S2 (ionization) signal~\cite{sorensen}. With this
method background rejection is less efficient, giving a significantly
weaker limit for $m_\chi \gtrsim 20$~GeV. However, due to the larger signal
in S2 compared to S1 the threshold can be lowered, allowing to put stronger
constraints on low-mass WIMPs independent of the $L_{\rm eff}$ ambiguities.
As illustrated in Fig.~\ref{fig:xenon} this analysis clearly excludes the
region relevant for CoGeNT/DAMA.

Given this situation it seems difficult to obtain a consistent
interpretation of all data in terms of SI interactions. This would require 
$(i)$ a major problem in the Na and/or Ge quenching factor determinations to
make DAMA and CoGeNT consistent, $(ii)$ a major calibration error in CDMS
(for Ge \cite{Ahmed:2010wy} and Si \cite{Akerib:2005kh} data), $(iii)$ a
major problem with the XENON S2-only analysis \cite{sorensen}, $(iv)$ very
low values of $L_{\rm eff}$ in Xenon in the $E_{nr} \sim$ few keV region.

\begin{figure}
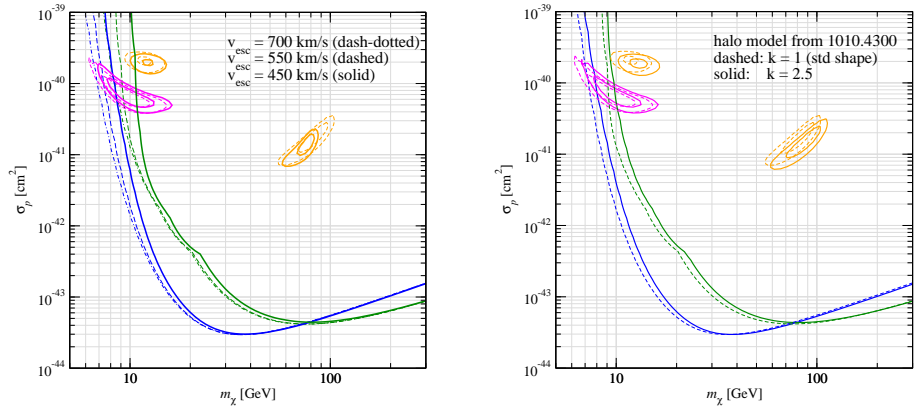

  \centering
  \includegraphics[width=0.37\textwidth]{halo-vesc.eps} \qquad
  \includegraphics[width=0.37\textwidth]{halo-shape.eps}
  \caption{DAMA (orange) and CoGeNT (magenta) regions at 90\% CL and
  $3\sigma$ compared to the 90\% CL exclusion curves from XENON10 (blue) and CDMS
  Ge 2009 (green). Left: the effect of changing the value of the galactic escape
  velocity. Right: changing the shape of the dark matter velocity
  distribution close to $v_{\rm esc}$ according to \cite{Lisanti:2010qx}. \label{fig:astro}}
\end{figure}

{\bf Can a modified dark matter halo reconcile the data?} The results obtained in
\cite{Fairbairn:2008gz} on previous data suggest that only quite extreem
(possibly unrealistic) assumptions on the dark matter velocity distribution
may lead to a slight improvement of the consistency of the data, for example
a highly anisotropic velocity distribution. In Fig.~\ref{fig:astro} we
illustrate---as two examples---effects related to the galactic escape
velocity. The left panel shows the effect of changing the value of the
escape velocity. Decreasing $v_{\rm esc}$ from 700 to 550~km/s can slightly
improve the consistency, since the allowed regions are practially unchanged
whereas the limits move slightly to larger $m_\chi$. In contrast, when
$v_{\rm esc}$ is lowered even further also the DAMA/CoGeNT regions start to
move to the right, and hence compatibility cannot be improved much further.
The default assumption adopted in all other figures shown here is $v_{\rm
esc} = 550$~km/s. In the right panel we investigate the impact of the shape
of the velocity distribution close to the cut-off as discussed recently in
\cite{Lisanti:2010qx}. Our standard assumption, a Maxwellian distribution
truncated at $v_{\rm esc}$ corresponds to the parameter $k = 1$ from
\cite{Lisanti:2010qx}, while $k=2.5$ describes a distribution going smoothly
to zero with a shape motivated by the dynamics of the outer part of the
galaxy. We observe that this modified distribution leads to a shift of the
regions as well as the constraints, without improving significantly the
relative compatibility. See also \cite{Fox:2010bz} on this topic.

{\bf Beyond elastic SI interactions.} Let us comment briefly on some
selected alternative particle physics models (without the ambition of being
complete). Assuming spin-dependent (SD) interactions it is possible to use
the fact that I and Na in DAMA have an odd number of protons, whereas
Xe and Ge have an even number of protons. Assuming that dark matter
interacts with the proton spin one can therefore evade the bounds for DAMA,
though no consistent explanation for CoGeNT is obtained in this way.
However, there are sever bounds from experiments using a florine target,
such as COUPP~\cite{Behnke:2008zza} or PICASSO~\cite{Archambault:2009sm},
which exclude the DAMA region at 90\%~CL~\cite{Kopp:2009qt}. Moreover,
in such a scenario neutrino constraints from the sun are tight
\cite{Desai:2004pq}, and under model dependent assumptions also
collider constraints rule out such a solution, e.g.~\cite{collider}.

If dark matter scatters inelastically to a slightly heavier dark particle,
annual modulation can be enhanced compared to the unmodulated rate, with a
different spectral shape, favouring heavy targets~\cite{TuckerSmith:2001hy}.
This scenario (assuming SI interactions) has been invoked to explain DAMA
(it cannot explain CoGeNT), but has been recently excluded by the W data
from CRESST~\cite{cresst}. A possible way out has been proposed
in~\cite{Kopp:2009qt} by assuming SD inelastic scattering on protons.
This provides a valid explanation of DAMA, avoiding constraints from Ge, Xe,
W, since these elements have an even number of protons, as well as
constraints from F experiments using the inelasticity to suppress scattering
on the light florine nucleus. Also in this scenario neutrino
constraints are important \cite{Shu:2010ta}, and in a given model one may
expect also collider constraints to be relevant due to the relatively large
cross section needed. 
A possible realization of SD inelastic interactions could originate from
tensorial interactions~\cite{Kopp:2009qt}. A related model is discussed in
\cite{Chang:2010en}, where dark matter interacts inelastically with nuclei via
the magnetic moment, using the exceptionally large magnetic moment of iodine
to explain the DAMA signal.

Finally, it is intriguing that DAMA as well as CoGeNT do not discriminate
between nuclear recoil and electronic events, whereas most other experiments
do. The assumption that dark matter interacts only with electrons but not
directly with quarks has been investigated in \cite{Kopp:2009et}, coming,
however, to a negative conclusion concerning a valid explanation of all
data.

{\bf Acknowledgements.} I would like to thank the IDM organizers for a very
pleasant and stimulating conference. I am grateful to J.~Kopp and J.~Zupan
for collaboration and lots of discussion on this topic. This work was partly
supported by the Transregio Sonderforschungsbereich TR27 ``Neutrinos and
Beyond'' der Deutschen Forschungsgemeinschaft.


\begin{thebibliography}{99}

\setlength{\itemsep}{-1mm}

\bibitem{dama}
  R.~Bernabei {\it et al.}  [DAMA Collaboration],
  Eur.\ Phys.\ J.\  C {\bf 56} (2008) 333
  [arXiv:0804.2741];\\
%
  P.~Belli, 
  these proceedings.

\bibitem{Bozorgnia:2010xy}
  N.~Bozorgnia, G.~B.~Gelmini and P.~Gondolo,
  JCAP {\bf 1011}, 019 (2010)
  [arXiv:1006.3110];\\
%
  G.~Gelmini, these proceedings.

\bibitem{cogent}
  C.~E.~Aalseth {\it et al.}  [CoGeNT collaboration],
  arXiv:1002.4703 [astro-ph.CO].

\bibitem{cresst}
  W.~Seidel, these proceedings.

\bibitem{cresst2}
  Talk by F.~Pr\"obst at ``Dark Matter: Direct Detection and Theoretical
  Developments'', Princeton, 15-16 November 2010,
  \verb!http://phy-gzk.princeton.edu/DMworkshop/Franz_Probst.pdf!

\bibitem{Hooper:2010uy}
  D.~Hooper, J.~I.~Collar, J.~Hall and D.~McKinsey,
  arXiv:1007.1005 [hep-ph].

\bibitem{Chagani:2008in}
  H.~Chagani {\it et al.}, 
  JINST {\bf 3}, P06003 (2008)
  [arXiv:0806.1916].

\bibitem{Ahmed:2009zw}
  Z.~Ahmed {\it et al.}  [The CDMS-II Collaboration],
  Science {\bf 327}, 1619 (2010)
  [arXiv:0912.3592];\\
%
  T.~Saab, these proceedings.

\bibitem{Akerib:2005kh}
  D.~S.~Akerib {\it et al.}  [CDMS Collaboration],
  Phys.\ Rev.\ Lett.\  {\bf 96}, 011302 (2006)
  [astro-ph/0509259].

\bibitem{Akerib:2010pv}
  D.~S.~Akerib {\it et al.}  [CDMS Collaboration],
  arXiv:1010.4290 [astro-ph.CO].

\bibitem{Ahmed:2010wy}
  Z.~Ahmed {\it et al.}  [CDMS-II Collaboration],
  arXiv:1011.2482 [astro-ph.CO].


\bibitem{xenon10}
  J.~Angle {\it et al.}  [XENON Collaboration],
  Phys.\ Rev.\ Lett.\  {\bf 100} (2008) 021303
  [arXiv:0706.0039];
%
  Phys.\ Rev.\  D {\bf 80}, 115005 (2009)
  [arXiv:0910.3698].

\bibitem{Aprile:2010um}
  E.~Aprile {\it et al.}  [XENON100 Collaboration],
  Phys.\ Rev.\ Lett.\  {\bf 105}, 131302 (2010)
  [arXiv:1005.0380];\\
%
  M.~Schumann, these proceedings.

\bibitem{savage}
  C.~Savage {\it et al.}, 
  arXiv:1006.0972 [astro-ph.CO];
%
  C.~Savage, these proceedings.

\bibitem{Bezrukov:2010qa}
  F.~Bezrukov, F.~Kahlhoefer and M.~Lindner,
  arXiv:1011.3990 [astro-ph.IM].

\bibitem{sorensen}
  P.~Sorensen, these proceedings.

\bibitem{Fairbairn:2008gz}
  M.~Fairbairn, T.~Schwetz,
  JCAP {\bf 0901}, 037 (2009)
  [arXiv:0808.0704].

\bibitem{Lisanti:2010qx}
  M.~Lisanti, L.~E.~Strigari, J.~G.~Wacker and R.~H.~Wechsler,
  arXiv:1010.4300 [astro-ph.CO].

\bibitem{Fox:2010bz}
  P.~J.~Fox, J.~Liu and N.~Weiner,
  arXiv:1011.1915 [hep-ph].

\bibitem{Behnke:2008zza}
  E.~Behnke {\it et al.} [COUPP Collaboration],
  Science {\bf 319 } (2008)  933-936
  [arXiv:0804.2886].


\bibitem{Archambault:2009sm}
  S.~Archambault {\it et al.} [PICASSO Coll.], 
  Phys.\ Lett.\  {\bf B682 } (2009)  185-192
  [arXiv:0907.0307].


\bibitem{Kopp:2009qt}
  J.~Kopp, T.~Schwetz, J.~Zupan,
  JCAP {\bf 1002 } (2010) 014
  [arXiv:0912.4264].

\bibitem{Desai:2004pq}
  S.~Desai {\it et al.} [Super-Kamiokande Coll.],
  Phys.\ Rev.\  {\bf D70 } (2004)  083523
  [hep-ex/0404025].

\bibitem{collider}
  J.~Goodman {\it et al.}, 
  arXiv:1005.1286;
%
  Y.~Bai, P.~J.~Fox, R.~Harnik,
  arXiv:1005.3797.

\bibitem{TuckerSmith:2001hy}
  D.~Tucker-Smith, N.~Weiner,
  Phys.\ Rev.\  {\bf D64 } (2001)  043502
  [hep-ph/0101138].

\bibitem{Shu:2010ta}
  J.~Shu, P.~Yin, S.~Zhu,
  Phys.\ Rev.\  {\bf D81 } (2010)  123519
  [arXiv:1001.1076].

\bibitem{Chang:2010en}
  S.~Chang, N.~Weiner, I.~Yavin,
  arXiv:1007.4200; N.~Weiner, these proceedings.

\bibitem{Kopp:2009et}
  J.~Kopp, V.~Niro, T.~Schwetz, J.~Zupan,
  Phys.\ Rev.\  {\bf D80}, 083502 (2009)
  [arXiv:0907.3159]; these proceedings.

\end{thebibliography}
\end{document}